\documentclass[conference]{IEEEtran}
\IEEEoverridecommandlockouts
\usepackage{amsthm}
\usepackage{float}
\usepackage{amsmath}
\usepackage{amssymb}
\usepackage{subfigure}
\usepackage{booktabs}
\usepackage{siunitx}
\usepackage{cite}
\usepackage{graphics, framed}
\usepackage{graphicx}
\usepackage{epsfig}
\usepackage{epstopdf}
\usepackage{url}
\usepackage{multimedia}
\usepackage{paralist}
\usepackage[printonlyused]{acronym}
\usepackage{units}
\usepackage[margin=0.91in]{geometry}
\usepackage{balance}
\usepackage[linesnumbered,ruled,vlined]{algorithm2e}

\renewcommand{\baselinestretch}{0.99}
\usepackage{color, soul}
\sethlcolor{yellow}
\makeatletter
\def\SOUL@hlpreamble{%
    \setul{\dp\strutbox}{\dimexpr\ht\strutbox+\dp\strutbox\relax}%
    \let\SOUL@stcolor\SOUL@hlcolor
    \SOUL@stpreamble
}

\makeatother

\newcommand{\FGR}[1]{Fig.~\ref{#1}}

\newcommand{\SEC}[1]{Section~\ref{#1}}
\newcommand{\TAB}[1]{Table~\ref{#1}}

\acrodef{SSA}[SSA]{signal and spectrum analyzer}
\acrodef{5G}[5G]{5\textsuperscript{th}--Generation}
\acrodef{FH}[FH]{frequency hopping}
\acrodef{CW}[CW]{continuous wave}
\acrodef{AWGN}[AWGN]{Additive White Gaussian Noise}
\acrodef{M2M}[M2M]{machine--to--machine}
\acrodef{dB}[dB]{decibel}
\acrodef{dBi}[dBi]{decibel isotropic}
\acrodef{dBm}[dBm]{decibel over a milliwatt}
\acrodef{DFT}[DFT]{discrete Fourier transform}
\acrodef{GHz}[GHz]{Gigahertz}
\acrodef{Hz}[Hz]{hertz}
\acrodef{IF}[IF]{intermediate frequency}
\acrodef{cfo}[CFO]{carrier frequency offset}
\acrodef{KHz}[KHz]{kilohertz}
\acrodef{PSD}[PSD]{power spectral density}
\acrodef{acf}[ACF]{autocorrelation function}
\acrodef{LOS}[LOS]{line--of--sight}
\acrodef{MHz}[MHz]{megahertz}
\acrodef{GHz}[GHz]{gigahertz}
\acrodef{RC}[RC]{radio controller}
\acrodef{STFT}[STFT]{short--time Fourier transform}
\acrodef{mmWave}[mmWave]{millimeter wave}
\acrodef{NGWN}[NGWN]{next generation wireless network}
\acrodef{NLOS}[NLOS]{non line--of--sight}
\acrodef{mse}[MSE]{mean square error}
\acrodef{OLOS}[OLOS]{optical line--of--sight}
\acrodef{PNA}[PNA]{performance network analyzer}
\acrodef{QoS}[QoS]{quality of service}
\acrodef{RF}[RF]{radio frequency}
\acrodef{spar}[s--parameter]{scattering parameters}
\acrodef{subThz}[sub--\unit{}{THz}]{sub--terahertz}
\acrodef{TUBITAK}[T\"{U}B\.{I}TAK]{Scientific and Technological Research Council of Turkey}
\acrodef{bilgem}[B\.{I}LGEM]{Informatics and Information Security Research Center}
\acrodef{USB}[USB]{universal serial bus}
\acrodef{VNA}[VNA]{vector network analyzer}
\acrodef{AoA}[AoA]{angle of arrival}
\acrodef{MLE}[MLE]{maximum likelihood estimation}
\acrodef{3D}[3D]{three dimensional}
\acrodef{2D}[2D]{two dimensional}
\acrodef{SNR}[SNR]{signal--to--noise ratio}
\acrodef{UAV}[UAV]{Unmanned aerial vehicle}
\acrodef{UAVa}[UAV]{unmanned aerial vehicle}
\acrodef{FHSS}[FHSS]{frequency hopping spread spectrum}
\acrodef{ISM}[ISM]{industrial, scientific, and medical}

\ifCLASSINFOpdf
\else
\fi

\begin{document}
\title{Measurement based FHSS--type Drone Controller Detection at 2.4GHz: An STFT Approach \thanks{This paper has been accepted for the presentation in the 2020 IEEE 91st Vehicular Technology Conference (VTC2020-Spring).}}
\IEEEoverridecommandlockouts 
\author{\IEEEauthorblockN{Batuhan Kaplan\IEEEauthorrefmark{1}\IEEEauthorrefmark{2}, \.{I}brahim Kahraman\IEEEauthorrefmark{1}\IEEEauthorrefmark{3}, Ali G\"{o}r\c{c}in\IEEEauthorrefmark{1}\IEEEauthorrefmark{4}, Hakan Ali \c{C}{{\i}}rpan\IEEEauthorrefmark{2}, Ali R{{\i}}za Ekti\IEEEauthorrefmark{1}\IEEEauthorrefmark{5}}
\IEEEauthorblockA{\IEEEauthorrefmark{1} Informatics and Information Security Research Center (B{\.{I}}LGEM), T{\"{U}}B{\.{I}}TAK, Kocaeli, Turkey}

\IEEEauthorblockA{\IEEEauthorrefmark{2} Department of Electronics and Communication Engineering, \.{I}stanbul Technical University, {\.{I}}stanbul, Turkey}

\IEEEauthorblockA{\IEEEauthorrefmark{3} Department of Electrical and Electronics Engineering, Bo\u{g}azi\c{c}i University, {\.{I}}stanbul, Turkey}

\IEEEauthorblockA{\IEEEauthorrefmark{4} Faculty of Electronics and Communications Engineering, Y{{\i}}ld{{\i}}z Technical University, {\.{I}}stanbul, Turkey}

\IEEEauthorblockA{\IEEEauthorrefmark{5} Department of Electrical and Electronics Engineering, Bal{{\i}}kesir University, Bal{{\i}}kesir, Turkey\\ Emails: \texttt{batuhan.kaplan@tubitak.gov.tr,} \texttt{ibrahim.kahraman@boun.edu.tr,}\\ \texttt{agorcin@yildiz.edu.tr,} \texttt{hakan.cirpan@itu.edu.tr,} \texttt{arekti@balikesir.edu.tr}}}

\maketitle

\begin{abstract}

The applications of the \acp{UAVa} increase rapidly in everyday life, thus detecting the \acp{UAVa} and/or its pilot is a crucial task. Many \acp{UAVa} adopt \ac{FHSS} technology to efficiently and securely communicate with their \acp{RC} where the signal follows a hopping pattern to prevent harmful interference. In order to realistically distinguish the \ac{FH} \ac{RC} signals, one should consider the real--world radio propagation environment since many \acp{UAVa} communicate with \acp{RC} from a far distance in which signal faces both slow and fast fading phenomenons. Therefore, in this study different from the literature, we consider a system that works under real--conditions by capturing over--the--air signals at hilly terrain suburban environments in the presence of foliages. We adopt the \ac{STFT} approach to capture the hopping sequence of each signal. Furthermore, time guards associated with each  hopping sequence are calculated using the \ac{acf} of the  \ac{STFT}  which  results  in  differentiating  the  each  \ac{UAVa} \ac{RC} signal accurately. In order to validate the performance of the proposed method, the results of normalized \ac{mse} respect to different \ac{SNR}, window size and Tx--Rx separation values are given.
\end{abstract}
\IEEEpeerreviewmaketitle
\acresetall
\section{Introduction}
\acp{UAV} have become a prevalent part of the daily life with their applications to many fields such as mapping and surveying, transportation, surveillance, law enforcement, aerial imaging and agriculture \cite{DHL_report}. Besides the aforementioned use of \acp{UAV} in many areas, one should keep in mind that \acp{UAV} can also be used dangerously to create unwanted incidents especially when they are diverted to the sensitive airspace near airports and their presence may cause accidents which can result in fatal crashes \cite{faa_report}. Moreover, \acp{UAV} can be utilized for collecting information about people, organisations, and companies without their consent. Therefore, identification of \ac{UAV} systems and their communication are great importance, especially to prevent unwanted situations. In this context, it is known that most of the communication between the \acp{UAV} and wireless \ac{RC} utilize the spread spectrum technology of \ac{FHSS} on \ac{ISM} band at \unit{2.4}{GHz} \cite{popovski}. Therefore a method to detect and classify these kinds of signals in this band would lead to the identification of the communication between the \ac{UAV} and the controller. 

In literature, most well known \ac{FHSS} signal detection methods adopt the time--frequency analysis and wavelet analysis \cite{kanaa2018robust,boashash2015time,wei2019robust,zhang2009blind}. It is shown that Wigner Ville distribution, Wavelet transform method, and array signal processing methods can be used to identify \ac{FHSS} signals with the burden of heavy computational complexity which make them hard to satisfy the real--time implementation. Since \ac{STFT} based detection method does not need any prior information about received signal compared to computationally complex algorithms, \ac{STFT} becomes the designated optimum detector in the absence of information regarding the received signal \cite{ouyang2001short}. However, one should note that in order to achieve good results with \ac{STFT}, \ac{SNR} value needs to be high. Thus, in this study, we propose \ac{STFT} based blind signal detection method for \ac{FHSS} \ac{UAV} \ac{RC} signals by using over--the--air signal measurements which accounts for the signal imperfections present in the nature of real--world conditions (e.g. fading, multipath, and so on) by considering the hilly terrain suburban environments in the presence of foliage. Furthermore, the literature utilizes the simulated data instead of over--the--air signals in general and these simulations assume that there is no time guards between hops. This assumption makes differentiation of \ac{FH} signals easier, however, many hopping signals use time guards and also these time guards are different for different signal sources. The proposed approach is also able to detect the time guards between each hopping sequence using the \ac{acf} \cite{chung1995parameter} of the \ac{STFT} which results in differentiating the each drone \ac{RC} signal accurately.


The paper is organized as follows, \SEC{sec:system_model} details the system model. \SEC{sec:proposed_method} presents the proposed method. The measurement setup is explained in \SEC{sec:measurement_setup}. In \SEC{sec:measurement_results}, measurement results are given and discussed. Finally, \SEC{sec:conclusion} concludes the paper.

\section{Signal Model}\label{sec:system_model}
Drone controller signals might be \ac{FH} signals that have temporal statistical characteristics and they can be written as \cite{wei2019robust},

\begin{equation}\label{eq:1}
x(t) = s(t) \sum_{m=0}^{M-1}  e^{j2 \pi f_{c_{m}} t_{m} + \theta_{m}} 
\end{equation}

\noindent where $\theta_{m}$ and $f_{c_{m}}$ are the carrier phase and carrier frequency of $m_{th}$ hop, respectively. Also, $s(t)$ denotes the complex baseband equivalent of the information bearer for $t \in [0,T]$, $M$ stands for the total number of hop of a signal, $t_{m}$ is the duration time of $m_{th}$ hop that might be a uniform distribution or not.

The received controller signal which is complex baseband equivalent of the received passband signal and can be expressed as,

\begin{equation}\label{eq:2}
    r(t) = \sum_{n=0}^{N-1} y_{n}(t) + n(t) + I(t)
\end{equation}

\noindent where $y_n(t)$ is an $n_{th}$ \ac{FH} signal source, $n(t)$ denotes the complex \ac{AWGN} in which I and Q components are i.i.d with $\mathcal{N}(0,\,\sigma^{2})$, $I(t)$ stands for the interference signal.

\subsection{Short--Time Fourier Transform}\label{sec:stft__model}

\ac{STFT} approach is utilized to analyze the \ac{FH} signals as a method to observe the frequency content of this type of non--stationary signals over time. Mathematical expression of \ac{STFT} of the time-domain signal $z(t)$ can be written as \cite{ouyang2001short},

\begin{equation}
    STFT\Big\{z(t)\Big\}=\int_{-\infty}^{\infty} [z(t) w(t-\tau)] e^{-j 2\pi f\tau} d \tau
\end{equation}

\noindent where $w(t)$ is the window function. The STFT matrix  $S = [s_1[f], s_2[f],  . . ., s_K[f]]$ such that $i_{th}$ element of this matrix is a column vector determined by discrete Fourier transform of $r[n] w[n-i R]$,

\begin{equation}\label{eq:stft}
    s_{i}[f]=\sum_{n=0}^{N-1} r[n] w[n-i R] e^{-j 2 \pi f n}
\end{equation}

\noindent where $r[n]$ is sampled version of $r(t)$ by considering the anti--aliasing property and $R$ denotes the shifting length.

One should keep in mind that adjusting the time and frequency resolution is crucial point for \ac{STFT} analysis \cite{kim2017comparison} due to the trade off between them. The length of the time point can be calculated as \cite{smith2011spectral}

\begin{equation}\label{msize}
    m=\left\lfloor\frac{N_{x}-L}{M-L}\right\rfloor
\end{equation}

\noindent where $N_x$ is length of the signal, $L$ denotes the number of overlap in the Fourier transform, $M$ represents the window size, $\lfloor \cdot \rfloor$ stands for the floor operator.
\section{Proposed Method}\label{sec:proposed_method}


The received controller signal, $r(t)$, analyzed by \ac{STFT} method. As depicted in \FGR{fig:flowgraph}, the flow graph explains how the system works in brief. After the signal is received in the first stage, optimal window time length is decided to get the optimum resolution at \eqref{msize} based on maximizing the number of elements on the matrix $S$ in the second stage of the flowchart. \ac{STFT} is calculated in a dBm unit according to \ac{PSD} in the same step.

\begin{figure}[!t]
    \centering
    \includegraphics[width=6cm, height=8cm]{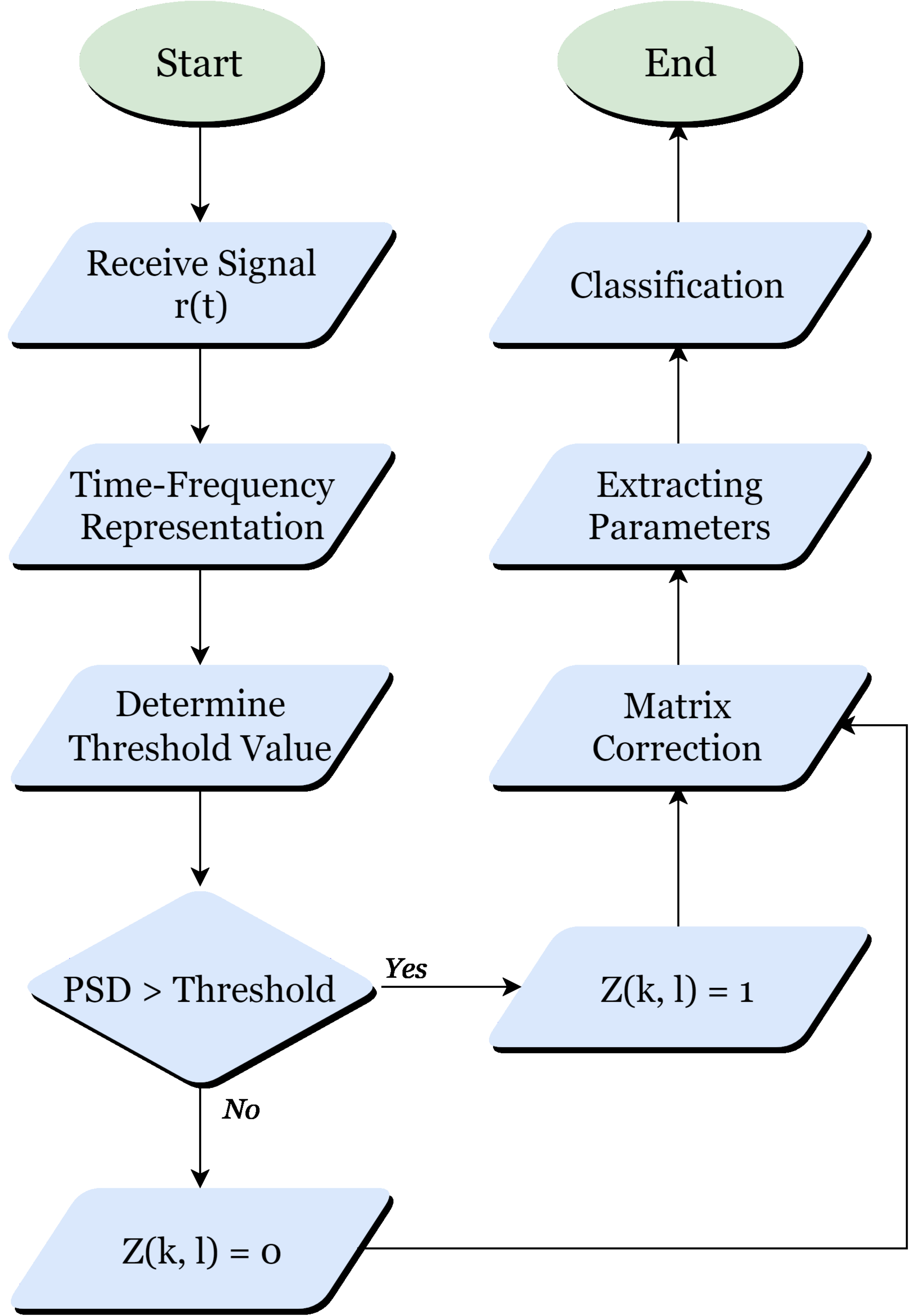}
    \caption{The flowchart of the proposed detection method.}
    \label{fig:flowgraph}
\end{figure}

A binarization process is conducted at the following steps of the flowchart; the \ac{STFT} matrix is converted to a binarized matrix, $Z(k, l)$, with the utilization of a threshold $\mu$. Also, in the flowchart \ac{PSD} refers to each point in the \ac{STFT} matrix. Based on the dynamically calculated threshold value, whether the signal presents or not is decided. Therefore, the problem statement can be indicated as the identification of the presence of the unknown \ac{FH} signal. When the signal is present, $Z(k, l)$ is evaluated as 1 and the new binarized matrix, $Z(k, l)$, is given as,

\begin{equation}
    Z(k, l) = 
    \begin{cases}
        1,  & S(k, l)\geq \mu\\
        0,  & S(k, l)< \mu
    \end{cases}
\end{equation}

To determine the threshold value, each element of \ac{STFT} matrix are concatenated and a sorting algorithm implemented to list the power levels of each point of the \ac{STFT} matrix in a ascending order. Then, we assume that the majority of the received signal comprise of noise, therefore, we take the mean value of the top $20\%$ of the sorted values of \ac{STFT} matrix to determine a lower bound for the computation of the threshold. Thus, the threshold is calculated as

\begin{equation}
    \mu = \frac{S_{max} + \sigma_{20\%}}{2}
\end{equation}

\noindent where $S_{max}$ is the maximum value of the STFT matrix, $\sigma_{20\%}$ denotes the mean of the top $20\%$ samples. \FGR{fig:thselect} shows $S_{max}$, $\sigma_{20\%}$, and the threshold value. Please note that even for the very low SNR regimes, the measurement results indicate the feasibility of this threshold selection process.

\begin{figure}[!t]
    \centering
    \includegraphics[width=\columnwidth]{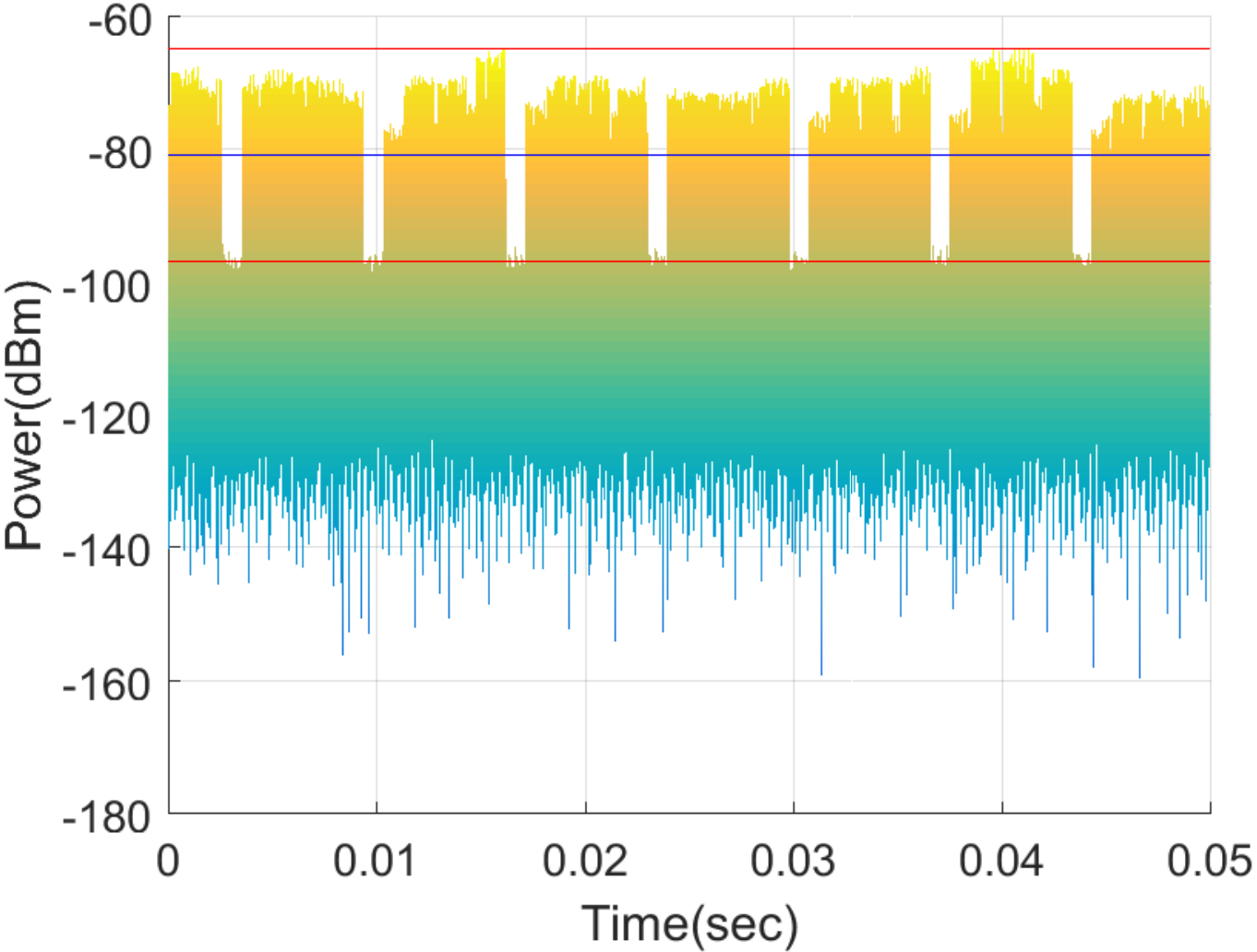}
    \caption{Estimated threshold representation with $S_{max}$ and $\sigma_{20\%}$ values over real recorded signal.}
    \label{fig:thselect}
\end{figure}

Due to the wireless impairments on the received signal, some portions of the $Z$ matrix might be missing as shown in \FGR{fig:sub1}. In order to represent the signal in a more plausible way, we adopt a widely used morphological dilation and erosion  processes from the domain of image processing \cite{haralick1987image,luo2009detection} to recover the received signal properly. A signal with impairments and the output of dilation and erosion processes are shown in \FGR{fig:dilation}.

\begin{figure*}[!t]
	\centering
	\subfigure[Recorded \ac{FH} drone controller signal with channel impairments.]{
		\label{fig:sub1}
		\includegraphics[width=.45\linewidth]{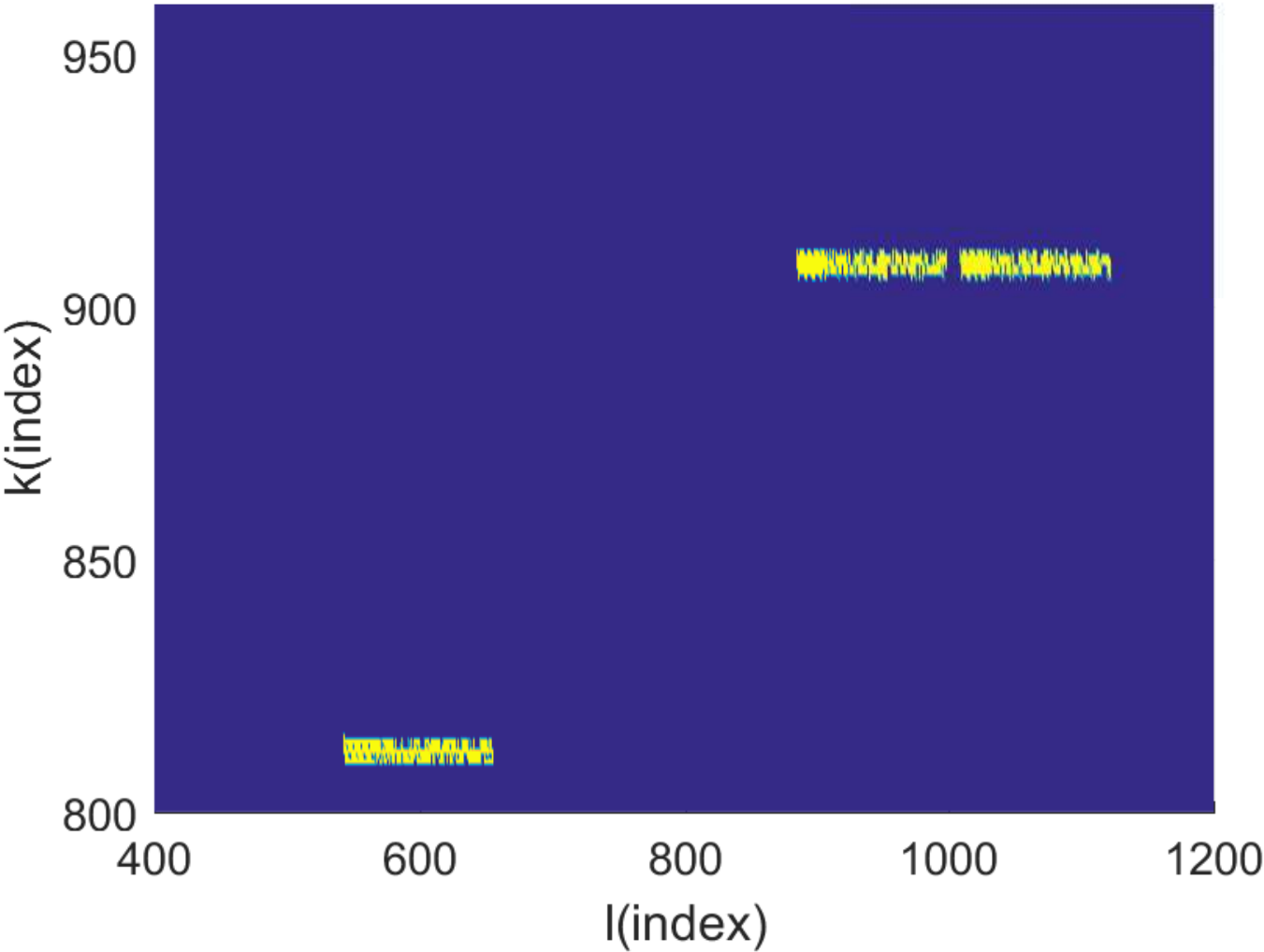}}
		\qquad
	\subfigure[\ac{FH} drone controller signal after correction.]{
		\label{fig:sub2}
		\includegraphics[width=.45\linewidth]{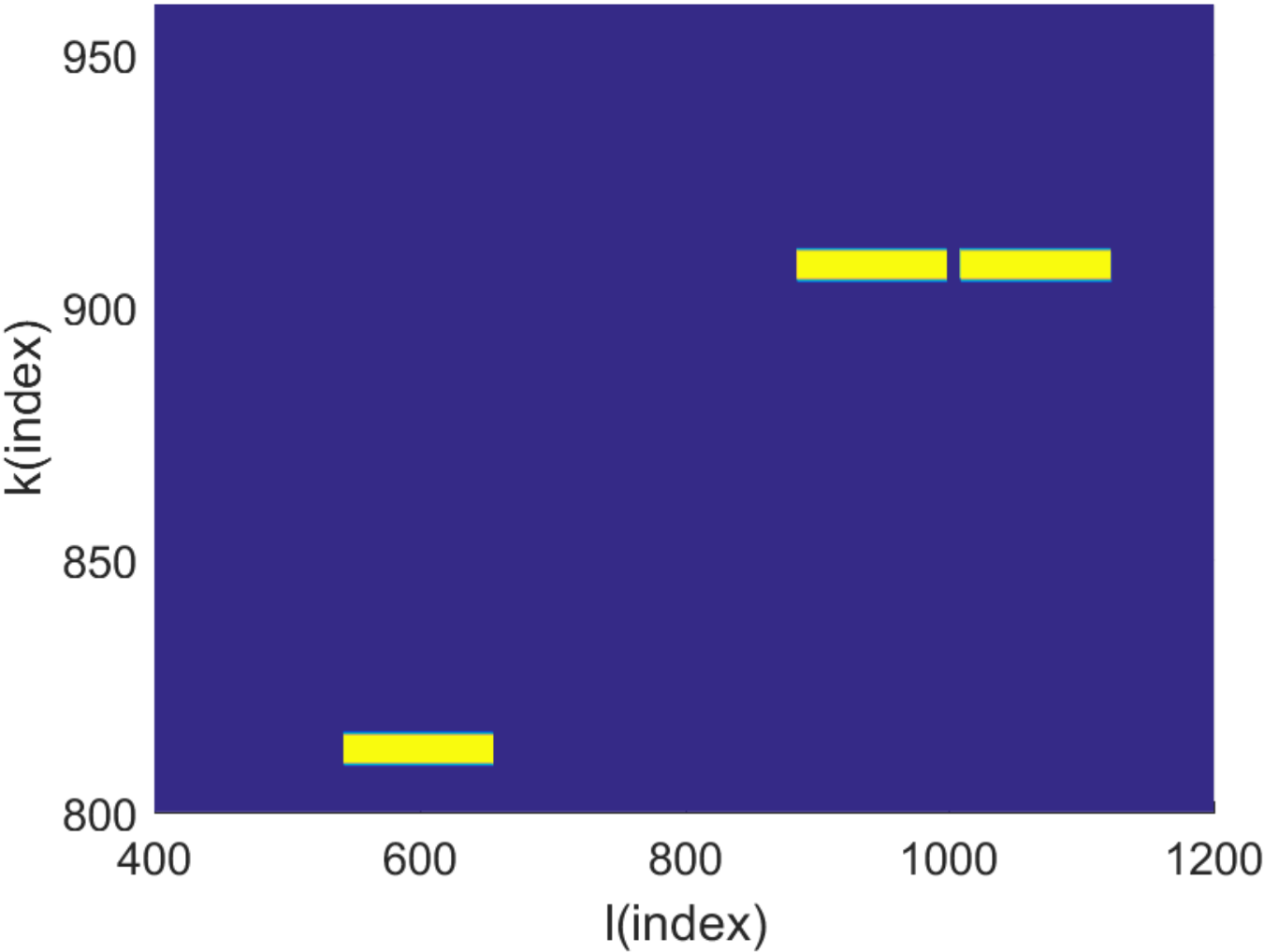}}
	\caption{Result of dilation process.}
	\label{fig:dilation}
\end{figure*}

After the recovery process, it becomes possible to extract the signal parameters such as start time, stop time, center frequency and difference between start time and stop time (dwell time) accurately. The parameter estimation algorithm is given in Algorithm \ref{algo:lineseg}. This process is implemented to the $Z$ matrix and the parameters of each signal in the spectrum are extracted. The algorithm simply detects the signals from $Z$ and extracts the duration of each independent transmission which can be part of same or different signal sources.

\begin{algorithm}[t]
	\SetKwInput{KwInput}{Input}
	\SetKwInput{KwOutput}{Output}
	\DontPrintSemicolon
	\label{algo:lineseg}
	\caption{Parameter Extraction Algorithm}
	\KwInput{$Z(k,l)$}
	\KwOutput{start time, stop time, center frequency, dwell time, bandwidth}
	
	\For{$i\leftarrow 1$ \KwTo \text{row}}{
	\For{$j\leftarrow 1$ \KwTo \text{column}}{
	$count \leftarrow 0$\\
	$bandwidth \leftarrow 0$
	
	\If{$Z(i,j)$ == \KwTo $1$}{
	\text{start time} $ \leftarrow j$ \\
	$f_{start} \leftarrow i$ \\
	\While{$Z(i,j)$ $==$ $1$ }{
	$count \leftarrow count+1$\\
	$j \leftarrow j+1$
	}
	\While{$Z(i,j-1)$ $==$ $1$ }{
	$bandwidth \leftarrow bandwidth+1$
	$i \leftarrow i+1$
	}
	\text{stop time} $\leftarrow j-1$ \\
	$f_{stop}\leftarrow i-1$ \\
	\text{center frequency} $\leftarrow \frac{f_{start}+f_{stop}}{2}$\\
	\text{dwell time} $\leftarrow count$ \\
	}
	\text{Assign $0$ to rectangle that founded above.}\\
	}
	}
\end{algorithm}


In the last step of the flowchart, controlling for each hop is done to decide whether it belongs to \ac{FH} signal or not. Due to the signal structure which will be explained in Section \ref{sec:measurement_setup}, time guards are inserted during hopping and this makes classification complicated. Time guards may lead to miss matches between hops because there can be some other signals between them and these miss matches should be corrected. In this paper \ac{acf} is utilized for this purpose since the highest correlation peaks occur as the signal matches itself perfectly and this is the case for drone controller FH signals based on the fact that time and frequency characteristics are fixed. \ac{acf} is used on $Z$ matrix and result of \ac{acf} gives the highest peak at $T_1$, which is the fundamental period (6.8ms) for the particular drone controller FH signal. The rest of the peaks give information in regards to hops which are generated from the same source. Thus, we define a set T = $\{T_1, T_2, ... T_n\}$ which represents the locations of all the peak values where $T_1 = \sup \hspace{1mm} \text{T}$. In other words, we put all the local extremum for interval $(0, T_1)$ to the set $\text{T}$ as shown in \FGR{fig:corr} meaning that we obtained all the required T values (guard and dwell times) to track and classify hops. Finally, the signal classification block in the flowchart is executed by the procedure: $\exists T_i \in \text{T}$ and any two hops, $hop\_a$ and $hop\_b$, are emitted from the same signal source if
\begin{equation}
    \text{start time($hop\_a$)} - \text{start time($hop\_b$)} \equiv T_i \hspace{1mm} (\text{mod } T_1)
\end{equation}

\noindent where start time is the estimated parameter from Algorithm \ref{algo:lineseg}. Thus, hop separation is also possible with the proposed method. Please note that even though the method is applied to a particular set of FH signals, it can be utilized to detect any kinds of FH signal with observable time and frequency domain hopping pattern. 

\begin{figure}[]
    \centering
    \includegraphics[width=\columnwidth]{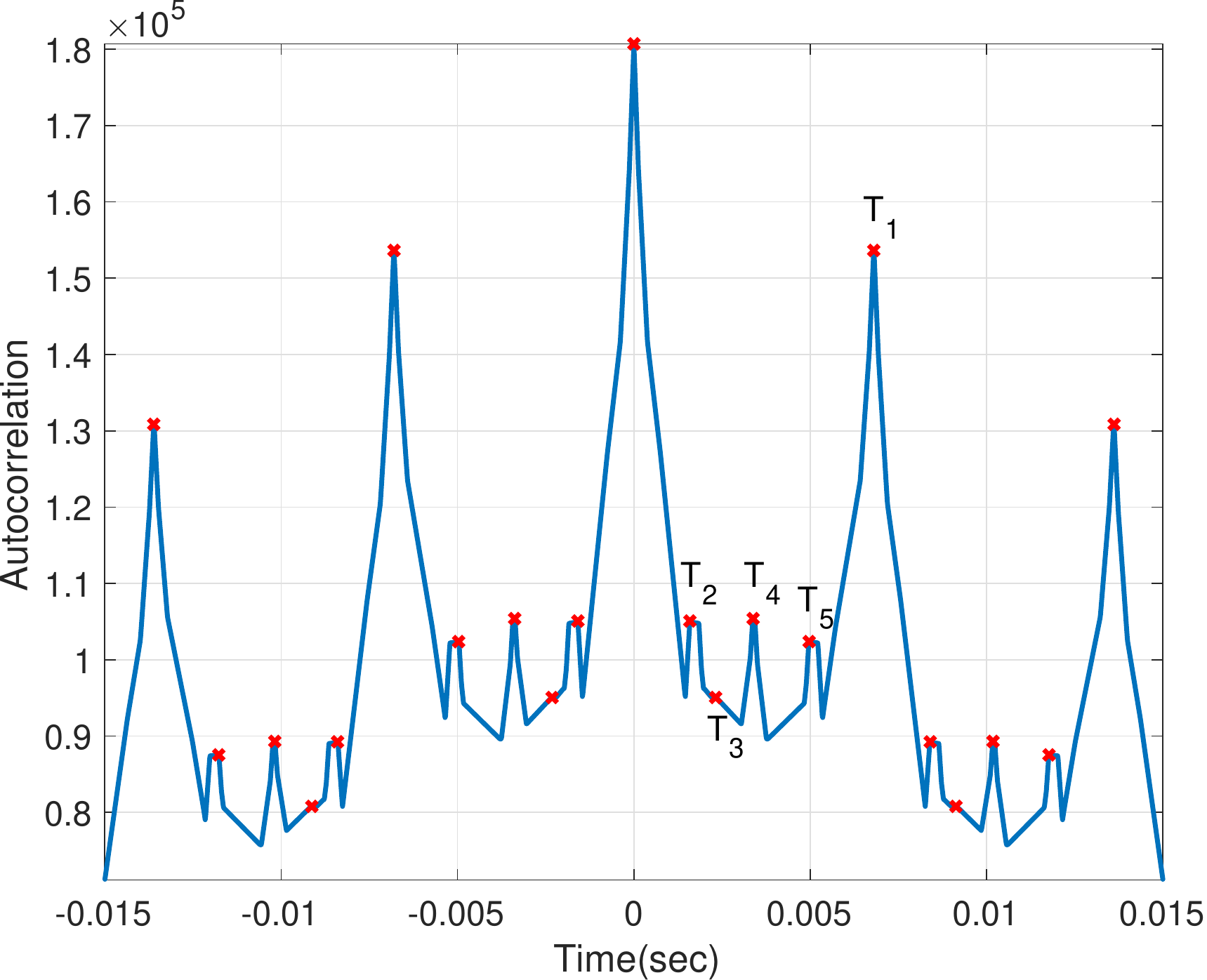}
    \caption{Time difference estimation between hops utilizing auto correlation function.}
    \label{fig:corr}
\end{figure}
\section{Measurement Setup}\label{sec:measurement_setup}
Experimental setup for \ac{FH} signal parameter estimation is performed in \ac{TUBITAK} \ac{bilgem}. Data collection of \ac{FH} signal is conducted with different distances from 5m to 135m with 10m intervals between each step. \FGR{fig:map} shows the fixed location of our receiver and locations of the drone controller (transmitter). 
\begin{figure}[!t]
    \centering
    \includegraphics[width=\linewidth,height=6.8cm]{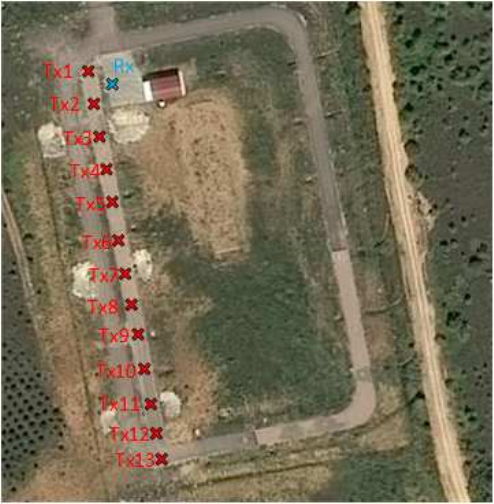}
    \caption{The aerial map of data collection locations.}
    \label{fig:map}
\end{figure}

\subsection{Hardware Setup}
The testbed used in the data acquisition procedure consists of signal source (drone controller) and spectrum analyzer to record the signals.

\begin{figure}[!t]
    \centering
    \includegraphics[width=\linewidth]{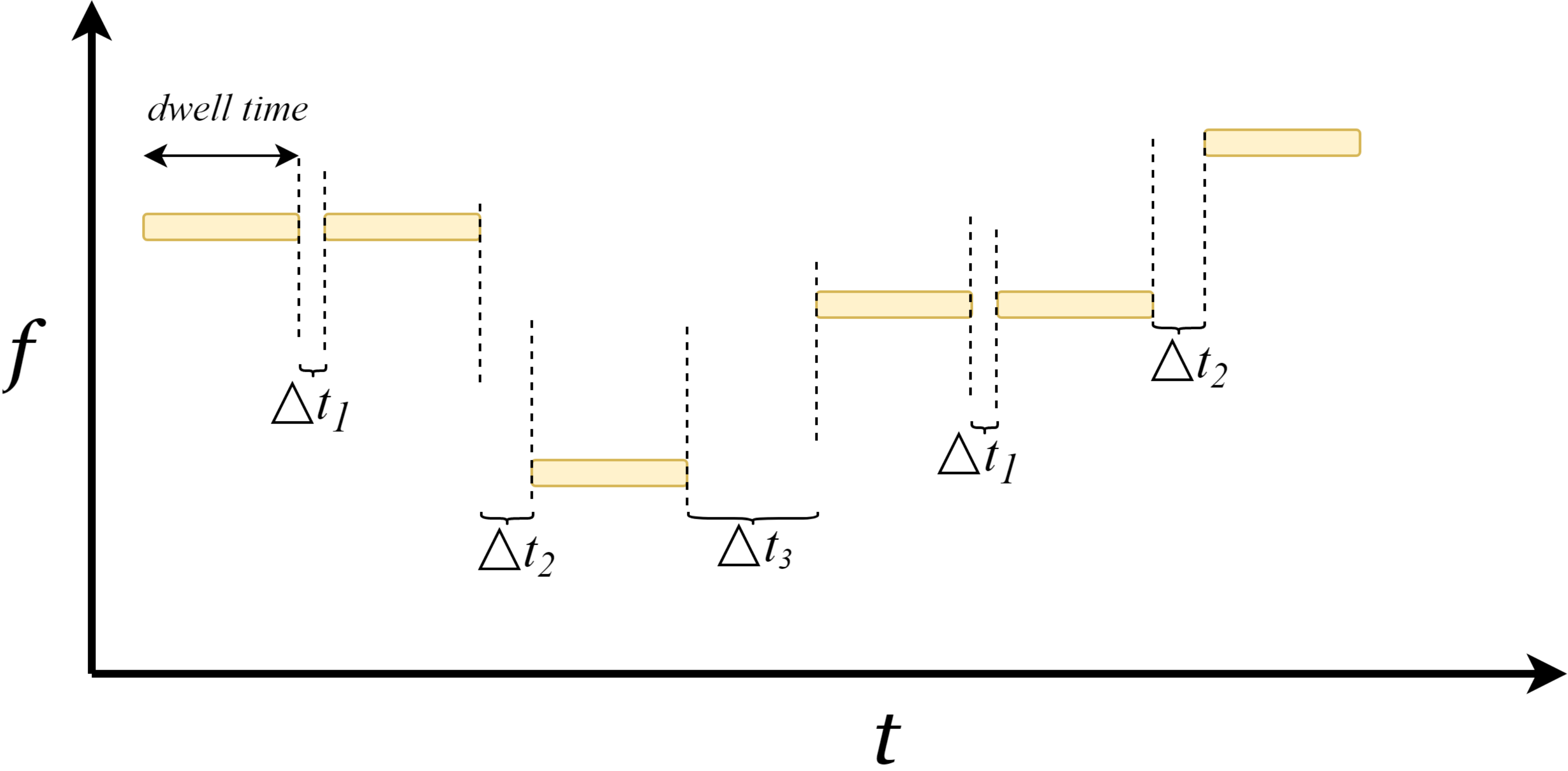}
    \caption{The hopping pattern of Drone controller signal: Futaba T8J \ac{RC} as signal source.}
    \label{fig:signalstr}
\end{figure}

Futaba T8J \ac{RC} is used during the experiment as a \ac{FH} signal source. Futaba T8J \ac{RC} operates in the 2.4 \ac{GHz} ISM band over the half of the spectrum band (up to 2.45 \ac{GHz}). When analyzing the signal, it is discovered that the \ac{RC} transmitter is behaved differently compared to the standard \ac{FH} communication systems (e.g. Bluetooth). An illustration of the periodic hopping sequence for the Futaba T8J \ac{RC} is shown in \FGR{fig:signalstr}.  Also, in the figure for $\tau_{dwell}$ represents dwell time and fundamental period becomes


\begin{equation}
    \Delta t_{1} < \Delta t_{2} < \Delta t_{3}
\end{equation}
\begin{equation}
    3\tau_{dwell}+\Delta t_{1} + \Delta t_{2} + \Delta t_{3}= T_1 =0.0068 sec
\end{equation}

\begin{table}[]
\centering
\caption{\ac{FH} signal characteristics }
\begin{tabular}{@{}ll@{}}
\toprule
\textit{\textbf{\ac{FH} Signal}} & \textit{\textbf{}} \\ \midrule
\textit{Dwell Time} & \textit{1.44ms} \\
\textit{Center Frequency Set} & \textit{\begin{tabular}[c]{@{}l@{}} 2.4GHz-2.45GHz interval\end{tabular}} \\
\textit{Hopping Sequence} & \textit{ f1      f1     f2     f3     f3     f4 ...} \\ \bottomrule
\end{tabular}
\label{tab:signalparam}
\end{table}  

The parameters of Futaba T8J \ac{RC} \ac{FH} signal source are listed in \TAB{tab:signalparam}. In the receiver side, Rohde\&Schwarz FSW 26 \ac{SSA} is utilized to record FH signals. \ac{SSA} can support the frequency range from 2 \ac{Hz} to 26.5\ac{GHz}. The device provides real--time spectral analysis up to 160 \ac{MHz} bandwidth. The signals are recorded over the 2.4\ac{GHz} \ac{ISM} spectrum band with an omnidirectional antenna.

\subsection{Experimental Procedures}
In data collection from real--world, it is assumed that transmitter operates between 2.4\ac{GHz}--2.48\ac{GHz} \ac{ISM} spectrum band. The center frequency of \ac{SSA} is set to 2.44\ac{GHz} and bandwidth of interest is adjusted to 80\ac{MHz} for the purpose of full coverage. Also, \ac{SSA} is connected to external computer via an Ethernet cable in favor of achieving data storage with ease. The sampling rate depends on the analysis bandwidth of real--time spectrum which is selected as 80MS/s. Each measurement is captured as an I/Q samples and collected 20M I/Q samples during 250ms. However, considering the processing limit, the collected data is divided into pieces with each has 4M samples. Each divided data was considered when calculating the performance of the proposed method. Finally, captured I/Q data is fed into the computer which runs MATLAB R2015b software. 

\section{Measurement Results}\label{sec:measurement_results}
Over--the--air data collection is realized and the performance of time--frequency analysis method is evaluated. Please note that all the captured data includes real--world propagation effects such as multipath fading, interference, \ac{cfo}. \FGR{fig:inputspec} shows how over–the–air recorded FH signal source behaves during the observation time. Please also note that some of the estimated parameters of the real signal which measured at 25m distance with 5dB \ac{SNR} can be found in \TAB{tab:parameters}.


\begin{figure}[!t]
    \centering
    \includegraphics[width=\linewidth]{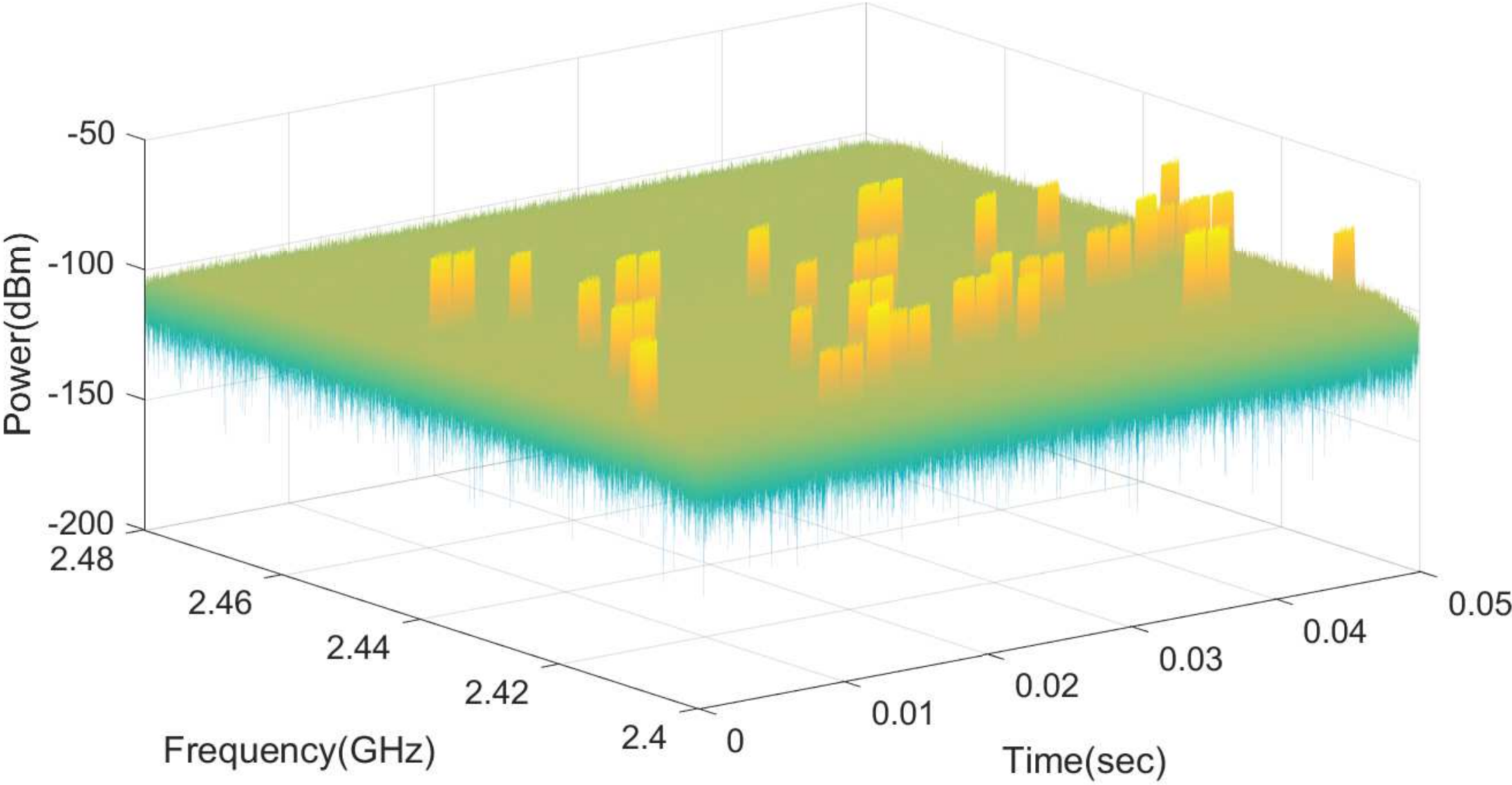}
    \caption{Recorded 2.4\ac{GHz} spectrum that comprised of drone controller FH signals.}
    \label{fig:inputspec}
\end{figure}

In order to validate the performance of the system, normalized \ac{mse} was considered. Normalized \ac{mse} can be calculated as\cite{ma2016blind}

\begin{equation}
    \text{NMSE} = \frac{1}{N}\sum_{i=1}^{N}\Big(\frac{\hat{t}_i-t}{t}\Big)^2
\end{equation}
\noindent where $\hat{t}_i$ is the estimated hopping time of $i_{th}$ hop and $t$ denotes the true value of hopping time. $N$ represents the total number of hop that must be found. When calculating the error, it is assumed that $N$ equals to $22$ in the observation time. Also,  $0$ values have been added to extracted parameters if there is some undetected signal.

\begin{table}[]
\centering
\caption{Estimated parameters of \ac{FH} signal}
\label{tab:parameters}
\begin{tabular}{@{}cccc@{}}
\toprule
\textit{\textbf{\begin{tabular}[c]{@{}c@{}}Start Time\\ (ms)\end{tabular}}} & \textit{\textbf{\begin{tabular}[c]{@{}c@{}}Stop Time\\ (ms)\end{tabular}}} & \textit{\textbf{\begin{tabular}[c]{@{}c@{}}Dwell Time\\ (ms)\end{tabular}}} & \textit{\textbf{\begin{tabular}[c]{@{}c@{}}Center Frequency\\ (GHz)\end{tabular}}} \\ \midrule
1.7930 & 3.2403 & 1.4472 & 2.4271 \\
5.1742 & 6.6086 & 1.4344 & 2.4414 \\
6.7623 & 8.1967 & 1.4344 & 2.4414 \\
8.6066 & 10.0410 & 1.4344 & 2.4211 \\
11.9749 & 13.4221 & 1.4472 & 2.4039 \\

\bottomrule
\end{tabular}
\end{table}

\begin{figure}[]
    \centering
    \includegraphics[width=\linewidth]{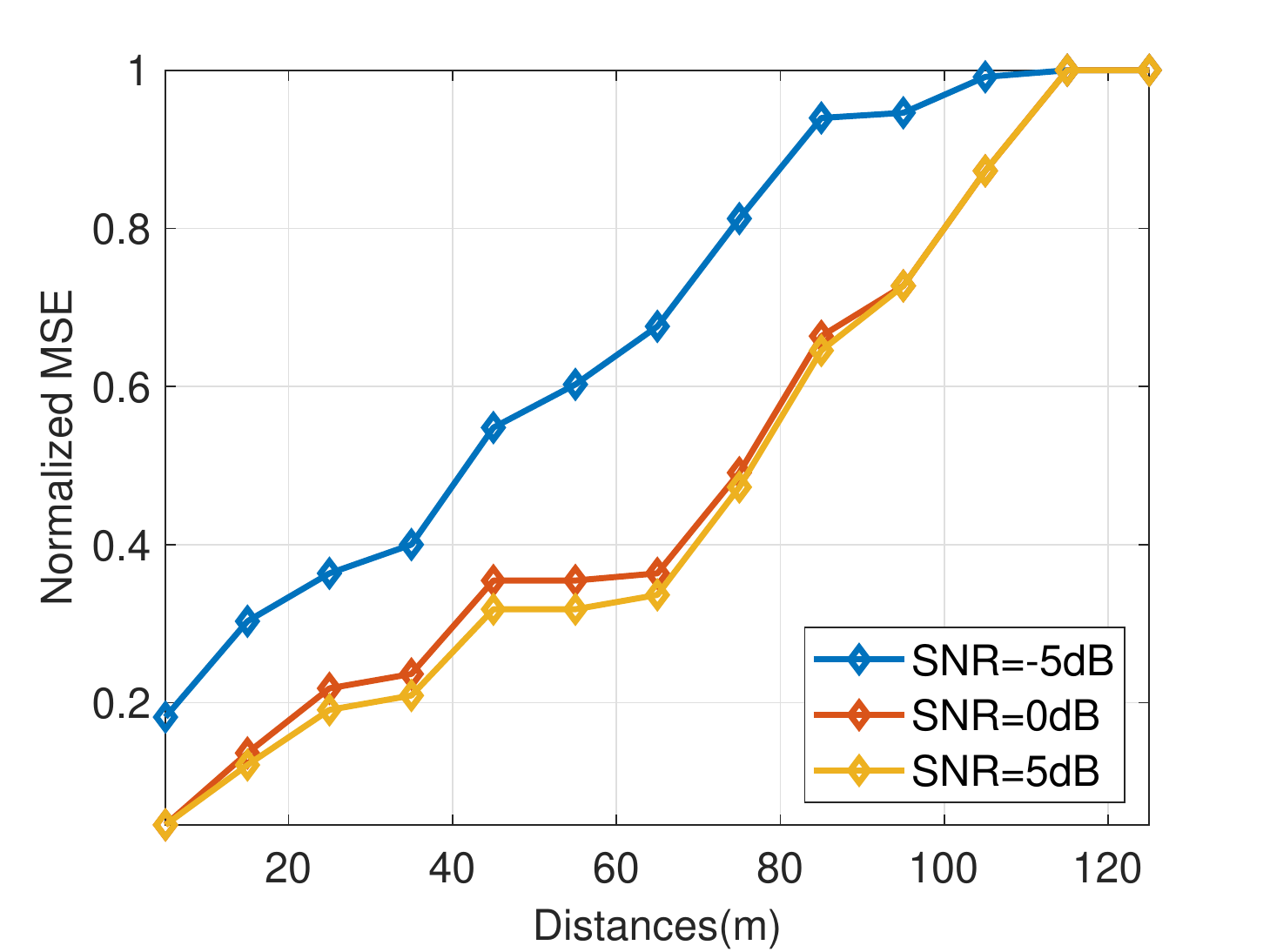}
    \caption{Estimated hopping time errors vs. distance in terms of normalized \ac{mse}.}
    \label{fig:mse}
\end{figure}

The error curve of an estimated hopping time from measured data is plotted in \FGR{fig:mse}. In here, error values are determined for three different \ac{SNR} values and distances. It is clearly seen that as the distance increases, the error of estimation increases. Also poor \ac{SNR} condition adversely affects the performance. Moreover, even with the same \ac{SNR} condition, there may be a missing hop of an \ac{FH} signal in the signal received from the farther point. 

As discussed before, another important issue is selection of window size ($M$). In this regard, precision of window size is also studied. It can be seen that window size directly effects the accuracy of the estimation. While window size decreases, we get high resolution in time but low resolution in frequency. In contrast, increasing window size implies high resolution in frequency but low resolution in time. Because both the frequency and time information are main features to classify the hopping signals, there should be some optimum window size providing better performance. It is empirically shown in \FGR{fig:window} that the optimum window size for STFT is $M=2048$. 

Since the previous works utilize \ac{STFT} only for deciding whether a signal is hopping between different frequencies, \ac{acf} is applied only in time--domain for the signals with no guard times, and the simulations were considered instead of measurements, comparison of the performance of the proposed method with these simulations are avoided in this work.


\begin{figure}[]
    \centering
    \includegraphics[width=\linewidth]{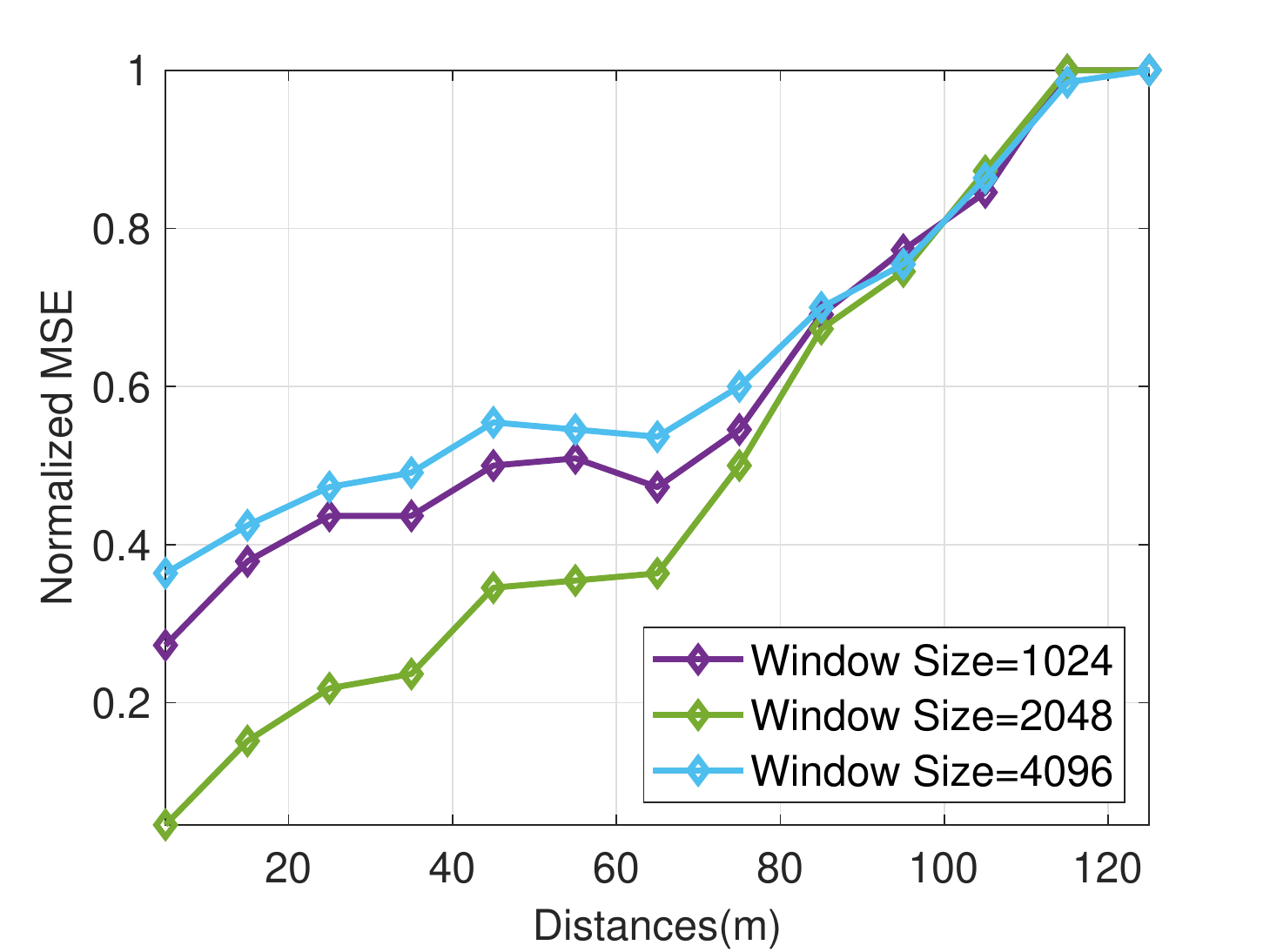}
    \caption{Estimated hopping time errors vs. distance for different window sizes and for \ac{SNR} $=$ 0 dB.}
    \label{fig:window}
\end{figure}

\section{Concluding Remarks and Future Directions}\label{sec:conclusion}
In this work, an FH drone controller signal detection algorithm is proposed and the performance of the algorithm is evaluated using measurements of \ac{UAV} \ac{RC} in real--world wireless environment. The algorithm can estimate the guards of the each hopping sequence successfully by using the \ac{acf} of \ac{STFT}, which also leads to the accurate identification whether a \ac{FH} signal is present or not. The performance of the proposed method is also quantified by normalized \ac{mse} for over--the--air signal which are recorded at different distances and SNRs. Measurement results show that reasonable input parameters will improve the performance of frequency hopping signal parameter estimation. In future studies, we will consider adopting the recently emerging deep learning algorithms to distinguish multiple standard based wireless hopping signals.
\section{Acknowledgement}
This publication was made possible by NPRP12S-0225-190152 from the Qatar National Research Fund (a member of The Qatar Foundation). The statements made herein are solely the responsibility of the author[s].
\balance
\bibliographystyle{IEEEtran}
\bibliography{VTC2020_Frequency_Hopping}

\begin{thebibliography}{10}
\providecommand{\url}[1]{#1}
\csname url@samestyle\endcsname
\providecommand{\newblock}{\relax}
\providecommand{\bibinfo}[2]{#2}
\providecommand{\BIBentrySTDinterwordspacing}{\spaceskip=0pt\relax}
\providecommand{\BIBentryALTinterwordstretchfactor}{4}
\providecommand{\BIBentryALTinterwordspacing}{\spaceskip=\fontdimen2\font plus
\BIBentryALTinterwordstretchfactor\fontdimen3\font minus
  \fontdimen4\font\relax}
\providecommand{\BIBforeignlanguage}[2]{{%
\expandafter\ifx\csname l@#1\endcsname\relax
\typeout{** WARNING: IEEEtran.bst: No hyphenation pattern has been}%
\typeout{** loaded for the language `#1'. Using the pattern for}%
\typeout{** the default language instead.}%
\else
\language=\csname l@#1\endcsname
\fi
#2}}
\providecommand{\BIBdecl}{\relax}
\BIBdecl

\bibitem{DHL_report}
DHL, ``{DHL launches first commercial drone 'parcelcopter' delivery service},''
  Available:
  \url{https://www.theguardian.com/technology/2014/sep/25/german-dhl-launches-first-commercial-drone-delivery-service},
  {Accessed: Oct. 29, 2019}.

\bibitem{faa_report}
FAA, ``{Federal Aviation Administration UAS Sightings Report},'' Available:
  \url{https://www.faa.gov/uas/resources/public_records/uas_sightings_report/},
  {Accessed: Oct. 29, 2019}.

\bibitem{popovski}
P.~{Popovski}, H.~{Yomo}, and R.~{Prasad}, ``Strategies for adaptive frequency
  hopping in the unlicensed bands,'' \emph{IEEE Wireless Communications},
  vol.~13, no.~6, pp. 60--67, Dec 2006.

\bibitem{kanaa2018robust}
A.~Kanaa and A.~Z. Sha’ameri, ``A robust parameter estimation of fhss signals
  using time--frequency analysis in a non-cooperative environment,''
  \emph{Physical Communication}, vol.~26, pp. 9--20, 2018.

\bibitem{boashash2015time}
B.~Boashash, \emph{Time-frequency signal analysis and processing: a
  comprehensive reference}.\hskip 1em plus 0.5em minus 0.4em\relax Academic
  Press, 2015.

\bibitem{wei2019robust}
S.~Wei, M.~Zhang, G.~Wang, X.~Sun, L.~Zhang, and D.~Chen, ``Robust multi-frame
  joint frequency hopping radar waveform parameters estimation under low
  signal-noise-ratio,'' \emph{IEEE Access}, 2019.

\bibitem{zhang2009blind}
X.~Zhang, X.~Wang, and X.-m. Du, ``Blind parameter estimation of
  frequency-hopping signals based on wavelet transform [j],'' \emph{Journal of
  Circuits and Systems}, vol.~4, 2009.

\bibitem{ouyang2001short}
X.~Ouyang and M.~G. Amin, ``Short-time fourier transform receiver for
  nonstationary interference excision in direct sequence spread spectrum
  communications,'' \emph{{IEEE} Trans. Signal Process.}, vol.~49, no.~4, pp.
  851--863, 2001.

\bibitem{chung1995parameter}
C.-D. Chung and A.~Polydoros, ``Parameter estimation of random fh signals using
  autocorrelation techniques,'' \emph{{IEEE} Trans. Commun.}, vol.~43, no.
  2/3/4, pp. 1097--1106, 1995.

\bibitem{kim2017comparison}
N.-K. Kim and S.-J. Oh, ``Comparison of methods for parameter estimation of
  frequency hopping signals,'' in \emph{International Conference on Information
  and Communication Technology Convergence (ICTC)}, 2017, pp. 567--569.

\bibitem{smith2011spectral}
\BIBentryALTinterwordspacing
J.~Smith, S.~U.~C. for Computer Research~in Music, Acoustics, and S.~U.~D.
  of~Music, \emph{Spectral Audio Signal Processing}.\hskip 1em plus 0.5em minus
  0.4em\relax W3K, 2011, online book, 2011 edition. [Online]. Available:
  \url{https://ccrma.stanford.edu/~jos/sasp/}
\BIBentrySTDinterwordspacing

\bibitem{haralick1987image}
R.~M. Haralick, S.~R. Sternberg, and X.~Zhuang, ``Image analysis using
  mathematical morphology,'' \emph{{IEEE} Trans. Pattern Anal. Mach. Intell.},
  no.~4, pp. 532--550, 1987.

\bibitem{luo2009detection}
L.~Luo \emph{et~al.}, ``Detection of an unknown frequency hopping signal based
  on image features,'' in \emph{2nd International Congress on Image and Signal
  Processing}, 2009, pp. 1--4.

\bibitem{ma2016blind}
Y.~Ma and Y.~Yan, ``Blind detection and parameter estimation of single
  frequency-hopping signal in complex electromagnetic environment,'' in
  \emph{2016 Sixth International Conference on Instrumentation \& Measurement,
  Computer, Communication and Control (IMCCC)}, 2016, pp. 370--374.

\end{thebibliography}
\end{document}